# The reference-frame independence of quantum probabilities

Benliang Li

*School of Mathematics and Physics, University of South China, Hengyang, 421001, China*

libenliang732@gmail.com

We investigate the transformation rule of a single particle wave-function under a change of reference frame. A postulate is raised, and some fundamental aspects regarding the reference-frame independence of quantum probabilities are explored.

## 1. Introduction

The probability which arises in the measurements of physical quantities plays a central role in the development of quantum physics. The wave-function of a single particle is a mathematical construct used to predict results of measurements, and its physical interpretation is given by the Born rule which attracts tremendous researches [1-6]. For a single quantum particle interacting with an external potential, the measurable physical quantities are spin, 4-momentum and the position of the particle, which are all encoded in its wave-function.

In a conference on *Foundations of Probability in Physics*, Kim raised a serious question: how would the probability distribution of a wave-function look to an observer in a different Lorentz frame [7]? And covariant wave functions was constructed to deal with the problem [8-9]. But a general wave-function describing one particle confined in an arbitrary potential can almost take any mathematical forms, and how these wave-functions observed in a different Lorentz frame still remains an open question.

In this paper, we explore the reference-frame independence of the quantum probabilities associated with particle's position measurements. We seek the solution to the basic and long-standing problem: how a wave-function describing a single particle moving in a potential transforms as observed in a different reference frame? We perform the transformations on the wave-function itself instead of the equations (such as Dirac equation or Schrödinger equation). Throughout this work we use natural units $c = \hbar = 1$.

## 2. Position Measurement of a free massless particle

Consider a free massless particle in one-dimension propagating towards Alice who sets a coordinate $(t, x)$. The wave-function is given by

$$\psi(t, x) = \int \phi(k) e^{i(\omega t - kx)} \mathrm{d}k \qquad (2.1)$$

in which $\omega = |k|$ is the energy of the particle, $\phi(k)$ stands for the normalized probability amplitude of measuring the momentum- $k$ and $\phi(k) = 0$ in case of $k < 0$, i.e., the modes of the wave all propagates in positive direction. Alice performs position measurement on the particle with the detector rest at position $x = 0$, and she sets $t = 0$ by the time the wave-front of the particle reaches position $x = 0$, as depicted by Fig. (1a).

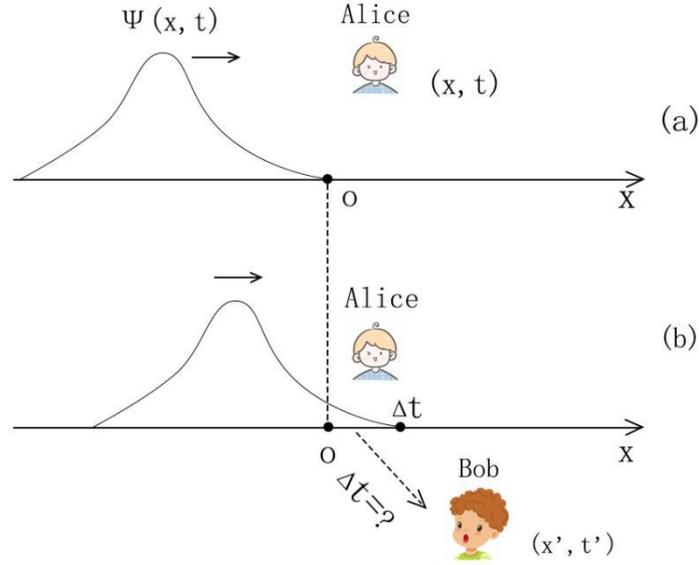

Fig. (1) A massless particle is propagating in positive direction, Alice is in an inertial frame $(x,t)$, Bob who is moving with constant velocity $v_B$ establishes another coordinate $(x',t')$. (a) depicts the wave at $t=0$; (b) depicts the wave at $t=\Delta t$, the detecting event occurred at $(t=\Delta t, x=0)$ is $(t'=\gamma\Delta t, x'=-\gamma v_B\Delta t)$ in Bob's frame, and the length of the wave passed through Bob is $\gamma\Delta t$, so the wave passed through the detector is $\gamma\Delta t + \gamma v_B\Delta t = \sqrt{(1+v_B)/(1-v_B)}\Delta t$.

Note that Eq. (2.1) is normalized both spatially and temporally, namely,

$$\int_{-\infty}^{\infty}\left|\psi(t,x)\right|^2\mathrm{d}x = \int_{-\infty}^{\infty}\left|\psi(t,x)\right|^2\mathrm{d}t = 2\pi\int_{-\infty}^{\infty}\left|\phi(k)\right|^2\mathrm{d}k = 1 \qquad (2.2)$$

This is due to the fact that the shape of the wave does not change during the propagation. Therefore, the detector can be fixed at $x=0$ during the whole measurement process, and it is accumulating the probability while the wave is passing through. The probability of measuring the particle at $x=0$ within infinitesimal time interval $(t, t+\Delta t)$ is given by

$$P(t,0) = \left|\psi(t,x=0)\right|^2\Delta t \qquad (2.3)$$

which can be regarded as the probability passing through the detector during the corresponding time interval.

To investigate the wave-function observed in a different reference frame, a *postulate* is raised:

*The outcomes of spatial measurements of a quantum particle are independent of the reference frames.*

This *postulate* is in analogy with the one in classical physics: when a physical object is found at space-time point $(t, x)$ in reference frame $S$, this same physical object must be found at a Lorentz transformed space-time point $(t', x')$ in frame $S'$. The difference is that the outcome at a specific space-time location is described as a probability which is generally less than 1 in quantum physics but is definite with probability 1 in classical physics.

To better illustrate the *postulate*, suppose Bob is moving with a constant velocity $v_B$ measured by Alice, he sets a coordinate $(t', x')$ in which the wave-function of the particle is written as $\psi'(t', x')$. The origin $(t' = 0, x' = 0)$ corresponds to $(t = 0, x = 0)$. Bob watched the whole process of the measurement performed by Alice, once Alice sees a detection event (the wave-function collapses) occurring at space-time point $(t, x = 0)$, Bob must observe the same detection event occurring at Lorentz transformed space-time point $(t' = \gamma t, x' = -\gamma v_B t)$ in which $\gamma = 1/\sqrt{1 - v_B^2}$. Here comes the crucial part: Alice can perform an infinite times of measurements and obtain the data matching the probability distribution of wave-function $\psi(t, x)$ as given by Eq. (2.1), then Bob who is watching all the measurements obtains the same data based on which he can construct $\psi'(t', x')$.

During the time interval $(t, t + \Delta t)$, the length of the wave passed through the detector is given by $\sqrt{\dfrac{1 + v_B}{1 - v_B}} \Delta t$ measured in coordinate $(t', x')$, as depicted by Fig. (1b), the probability can be given as

$$P'(t', x') = |\psi'(t', x')|^2 \sqrt{\frac{1 + v_B}{1 - v_B}} \Delta t \qquad (2.4)$$

which is the probability passing through the detector during time interval $(t, t + \Delta t)$ observed by Bob. Since Alice and Bob share the same data from Alice's measurement outcomes, the probability of measuring the particle at $x = 0$ within time interval $(t, t + \Delta t)$ viewed by Alice, is the same as the probability given by Eq. (2.4), namely, the quantum probability is invariant in different Lorentz frames. Therefore, it is required that

$$\frac{P(t_1, 0)}{P(t_2, 0)} = \frac{P'(t_1', x_1')}{P'(t_2', x_2')} \qquad (2.5)$$

in which $t_1$ and $t_2$ are two arbitrary time instants. $(t_1', x_1')$ and $(t_2', x_2')$ are Lorentz transformed of $(t_1, 0)$ and $(t_2, 0)$, respectively. Since Eq. (2.5) holds for any values of $t_1$ and $t_2$, we demand

$$\psi'(t', x') = A\psi(t, x) \qquad (2.6)$$

in which $(t', x')$ is Lorentz transformed of $(t, x)$, $A$ is the normalization constant satisfying

$$\int_{-\infty}^{\infty} |\psi'(t', x')|^2 \, dt' = 1 \qquad (2.7)$$

Note that the integration in Eq. (2.7) is performed along Bob's path, i.e.,

$\mathrm{d}x' = 0$, then we have $\mathrm{d}t' = \gamma(\mathrm{d}t - v_B \mathrm{d}x) = \mathrm{d}t / \gamma$.

For free massless particle given by Eq. (2.1), the Lorentz transformed wave-function can be given as

$$\psi'(t', x') = \int \phi'(k') e^{i(\omega' t' - k' x')} \mathrm{d}k' \qquad (2.8)$$

in which $\begin{cases} \omega' = \gamma(\omega - v_B k) \\ k' = \gamma(k - v_B \omega) \end{cases}$ is Lorentz transformed energy-momentum

vector $(\omega, k)$. We also have $\phi'(k') = \phi(k)$, one can check

$\psi'(t', x') = \sqrt{\dfrac{1 - v_B}{1 + v_B}} \psi(t, x)$ which satisfies Eq. (2.6). Thus, Eq. (2.6) together

with $\phi'(k') = \phi(k)$, implies that the wave-function is a scalar under Lorentz transformation both in the real space $(t, x)$ and in the energy-momentum space $(\omega, k)$, and Lorentz transformation on wave-functions only shifts the probability distribution to the transformed point without distorting it. Note that this conclusion applies for free massless particles with a characteristic that the shape of the wave is not changing during propagation, i.e., the wave is non-dispersive.

In general, a massive quantum particle interacting with an external potential is off-shell, the wave-function viewed in $(t, x)$ frame can be given as

$$\psi(t, x) = \int \phi(k) \varepsilon(\omega) e^{ikx} e^{-i\omega t} \mathrm{d}k \mathrm{d}\omega \qquad (2.9)$$

in which $\phi(k)$ and $\varepsilon(\omega)$ are two unrelated functions representing the probability distribution in $k$-space and $\omega$-space, respectively. Especially for a wave-function with eigen-energy $\omega_E$, we have

$\varepsilon(\omega) = \delta(\omega_E)$, Eq. (2.9) becomes

$$\psi_{\omega_E}(t,x) = e^{-i\omega_E t} \int \phi(k) e^{ikx} \mathrm{d}k \qquad (2.10)$$

The mathematical forms of Eqs. (2.9) and (2.10) depend on the potential which can take arbitrary mathematical expressions since limitless configurations of sources can be constructed [10]. Eq. (2.9) does not satisfy Eq. (2.2) since the shape of the wave is changing over time. Then whether the analysis, based on which to derive Eq. (2.6), can be applied for a massive particle in an external potential is in doubt. Indeed, how Eq. (2.9) transforms by changing Lorentz frames is still unclear, whether there exists a general transformation rule which can be applied to the wave-function regardless of the form of the potentials interacting with the particle is still an open question. In the rest of this paper, we develop a novel strategy to solve this problem.

## 3. Position measurement of a non-free particle

In reality, the act of quantum measurement takes some finite time duration to complete; therefore, we introduce $\Delta t_d$ as the time duration of a single detection process. To illustrate the meaning of $\Delta t_d$, a negatively-charged particle is confined in a 2-dimensional rectangular box with length $a$ and width $b$, as depicted in Fig. (2). In order to measure the position distribution of the particle, electrons (or other detecting particles) at different locations are projected into the box, the wave-function of the confined particle collapses to a state with a certain position once the electron at the corresponding location is reflected from the impact. Note that the detecting particles need to be prepared with the same initial velocity $v_d$ along $x$ direction, then we have $\Delta t_d = b / v_d$ which is independent of $x$. Note that $\Delta t_d$ can be an infinitesimal value in case of $b$ equals to Bohr radius and $v_d = 1$.

**detecting particles**

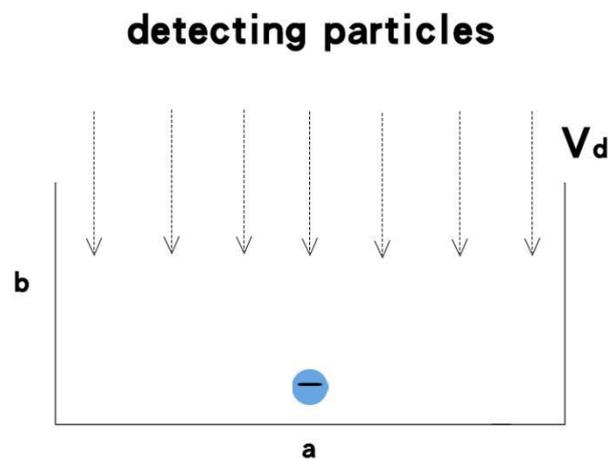

Fig. (2) A negatively charged particle is confined in a box, detecting particles with the same velocity $v_d$ are projected into the box to measure the position of the charged particle.

We focus on the description of a single particle confined in a 1-dimensional quantum well with spatial extension $x \in [0, L]$, and the wave-function of the particle can be written as $\psi(t, x)$.

Alice sets a coordinate $(t, x)$ and prepares $N$ identical copies of wave-functions $\psi(t, x)$ as depicted in Fig. (3). Each wave-function describes 1 charged particle confined in a 1-dimensional quantum well with finite space extension $[0, L]$ measured in coordinate $(t, x)$, the quantum well can be non-static which produces a time-dependent wave-function $\psi(t, x)$. At time $t = 0$, Alice performs the position measurements on every sample simultaneously, and for each measurement the particle reveals its position somewhere in its corresponding quantum well. The time duration of one detection process denoted as $\Delta t_d$ is also recorded. Once all the measurements are completed at time $t = \Delta t_d$, Alice prepares another $N$ identical samples and repeats the measurements; such procedure is repeated $m$ times in succession with time intervals of $\Delta t_d$, the data of the measurement outcomes are shown by Table (1). In reality, a particle cannot occupy a single point in space but requires a certain extension which is denoted as $\Delta x$, as a result, the space is cut into $L / \Delta x$ slices. Within time interval $[i\Delta t_d, (i+1)\Delta t_d]$ and space interval $[j\Delta x, (j+1)\Delta x]$ in which $j$ are integers with condition $0 \le j < L / \Delta x$, the particle is measured $^i n_j$ times satisfying

$$^i n_0 + {^i n_1} + \cdots + {^i n_j} + \cdots + {^i n_{(L - \Delta x)/\Delta x}} = N \qquad (3.1)$$

for any $i$-th row in Table (1).

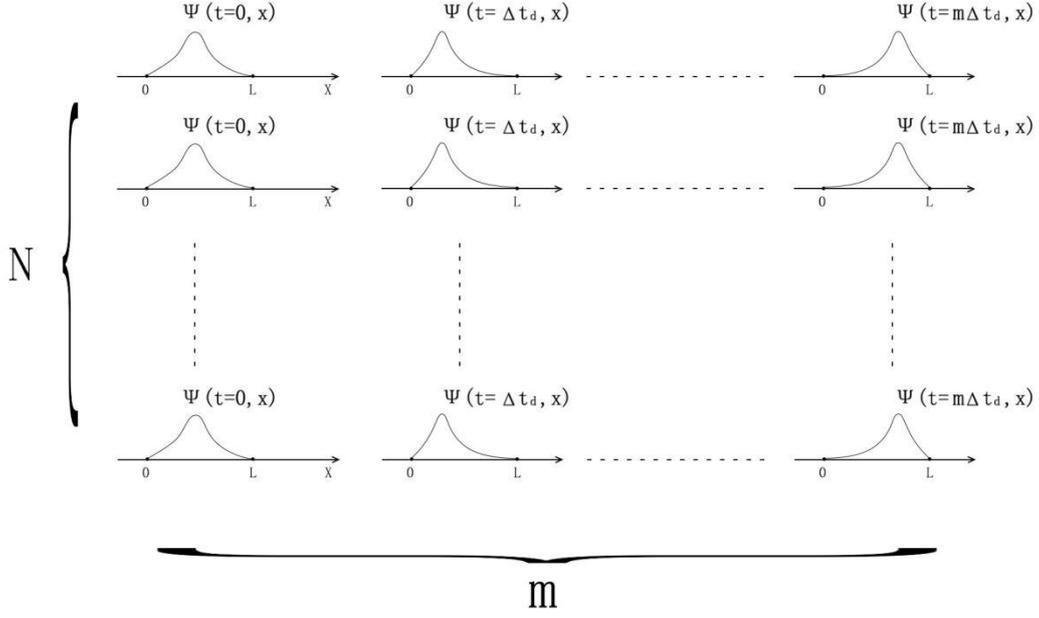

Fig. (3) N×m measurements are conducted in succession, each time the measurements are performed on N identical samples.

For time-dependent wave-function $\psi(t,x)$, the outcomes in $i$-th row represent the probability distribution of the particle at time $t = i\Delta t_d$ which corresponds to the wave-function $\psi(i\Delta t_d, x)$. For the quantum tomography process performed by Alice, $N$ needs to be a big number in order to reconstruct $\psi(t,x)$ precisely from the data shown by Table (1). Note that within the infinitesimal time period $\Delta t_d$ of the detecting process, the wave-function can be treated as static, namely,

$$\psi(i\Delta t_d \leq t < i\Delta t_d + \Delta t_d, x) = \psi(i\Delta t_d, x) \qquad (3.2)$$

for any values of $x$ and $i$. Then the probability corresponding to the data in $i$-th row of Table (1) is given as

$$P(i\Delta t_d, j\Delta x) = \left|\psi(i\Delta t_d, j\Delta x)\right|^2 \Delta x = \frac{{}^i n_j}{N} \qquad (3.3)$$

with normalization condition $\int_0^L |\psi(t,x)|^2 \mathrm{d}x = 1$ for any time $t$.

| $t=0$ | $^0n_0$ | $^0n_1$ | $^0n_2$ | $^0n_3$ | $^0n_4$ | …… | $^0n_{j-1}$ | $^0n_j$ | $^0n_{j+1}$ | …… | $^0n_{(L-\Delta x)/\Delta x}$ |
|---|---|---|---|---|---|---|---|---|---|---|---|
| $t=\Delta t_d$ | $^1n_0$ | $^1n_1$ | $^1n_2$ | $^1n_3$ | $^1n_4$ | …… | $^1n_{j-1}$ | $^1n_j$ | $^1n_{j+1}$ | …… | $^1n_{(L-\Delta x)/\Delta x}$ |
| $t=2\Delta t_d$ | $^2n_0$ | $^2n_1$ | $^2n_2$ | $^2n_3$ | $^2n_4$ | …… | $^2n_{j-1}$ | $^2n_j$ | $^2n_{j+1}$ | …… | $^2n_{(L-\Delta x)/\Delta x}$ |
| $t=3\Delta t_d$ | $^3n_0$ | $^3n_1$ | $^3n_2$ | $^3n_3$ | $^3n_4$ | …… | $^3n_{j-1}$ | $^3n_j$ | $^3n_{j+1}$ | …… | $^3n_{(L-\Delta x)/\Delta x}$ |
| $\vdots$ $\vdots$ | | $\vdots$ $\vdots$ | | $\vdots$ $\vdots$ | | $\vdots$ $\vdots$ | | $\vdots$ $\vdots$ | | $\vdots$ $\vdots$ | |
| $t=i\Delta t_d$ | $^in_0$ | $^in_1$ | $^in_2$ | $^in_3$ | $^in_4$ | …… | $^in_{j-1}$ | $^in_j$ | $^in_{j+1}$ | …… | $^in_{(L-\Delta x)/\Delta x}$ |
| | | | | | | | | | | | |
| $t=m\Delta t_d$ | $^mn_0$ | $^mn_1$ | $^mn_2$ | $^mn_3$ | $^mn_4$ | …… | $^mn_{j-1}$ | $^mn_j$ | $^mn_{j+1}$ | | $^in_{(L-\Delta x)/\Delta x}$ |

Table (1)

Bob is moving with a constant velocity $v_B$ measured by Alice, he sets a coordinate $(t',x')$ in which the wave-function of the particle is written as $\psi'(t',x')$. The origin $(t'=0,x'=0)$ corresponds to $(t=0,x=0)$. Bob watched the whole process of $m \times N$ measurements performed by Alice, once Alice sees a detection event occurring within space interval $x \in (j\Delta x, j\Delta x + \Delta x)$ at time $t=i\Delta t_d$, Bob must observe the same event occurring on a straight line within two Lorentz-transformed space-time points given by $\begin{cases} t' = \gamma(i\Delta t_d - v_B j\Delta x) \\ x' = \gamma(j\Delta x - v_B i\Delta t_d) \end{cases}$ and $\begin{cases} t' = \gamma[i\Delta t_d - v_B(j+1)\Delta x] \\ x' = \gamma[(j+1)\Delta x - v_B i\Delta t_d] \end{cases}$ in which

$\gamma = 1/\sqrt{1 - v_B^2}$ .

We acknowledge that the measurements performed at $(t, x)$ with $t = v_B x$ are simultaneous events observed in Bob's reference frame $(t', x')$. As shown in Table (1), different velocities $v_B$ correspond to straight lines with different slopes, and the measurement outcomes crossed by the corresponding straight lines can be treated as events occurred simultaneously in frame $(t', x')$. For instance, $v_B = \Delta t_d / \Delta x$ is shown by the orange line, and the measurement outcomes $({}^0n_0, {}^1n_1, {}^2n_2, \cdots, {}^kn_k, \cdots)$ on the orange line can be treated as events occurred simultaneously in frame $(t', x')$. Therefore, these outcomes correspond to the probability distribution of the wave-function $\psi'(t', x')$ at $t' = 0$. Similarly, measurement outcomes $({}^0n_0, {}^0n_1, {}^1n_2, {}^1n_3, {}^2n_4 \cdots)$ on the middle blue line which represents $v_B = \Delta t_d / (2\Delta x)$ correspond to the probability distribution of the wave-function $\psi'(t', x')$ at $t' = 0$ ; $({}^0n_2, {}^0n_3, {}^1n_4, {}^1n_5, \cdots)$ on the upper blue line correspond to the probability distribution at $t' = -\gamma \Delta t_d$ and $({}^1n_0, {}^1n_1, {}^2n_2, {}^2n_3, \cdots)$ on the lower blue line correspond to the probability distribution at $t' = \gamma \Delta t_d$.

Suppose two detection events occurred at $(i\Delta t_d, j\Delta x)$ and $(i\Delta t_d + \Delta t_d, j\Delta x + \Delta x)$ are simultaneous events observed by Bob, then the spatial distance between these two events is $\Delta x' = \Delta x / \gamma$, i.e., the detection length $\Delta x$ in Eq. (3.3) becomes $\Delta x'$ in Bob's frame, this is the length contraction effect in special relativity. For $v_B = \dfrac{i\Delta t_d}{j\Delta x}$, the data ${}^0n_0$ and ${}^in_j$ are measured simultaneously observed by Bob, we have

$$\frac{P'(0, j\Delta x')}{P'(0,0)} = \frac{\left|\psi'(t'=0, x'=j\Delta x')\right|^2 \Delta x'}{\left|\psi'(t'=0, x'=0)\right|^2 \Delta x'} = \frac{^i n_j}{^0 n_0} \tag{3.4}$$

We already know that

$$\frac{P(i\Delta t_d, j\Delta x)}{P(0,0)} = \frac{\left|\psi(i\Delta t_d, j\Delta x)\right|^2 \Delta x}{\left|\psi(0,0)\right|^2 \Delta x} = \frac{^i n_j}{^0 n_0} \tag{3.5}$$

from Eq. (3.3), apply the transformation rule given by Eq. (2.6) to this case, Eq. (3.4) is satisfied. Since the the data $^0 n_0$ and $^i n_j$ in Eqs. (3.4) and (3.5) are arbitrary, therefore, Eq. (2.6) can be applied in general for wave-functions of non-free massive particles.

Note that, in general, the summation of the data on the corresponding line in Table (1) do not equal to $N$, only the ratio between these data represents the probability distribution of the wave-function viewed by Bob. To be specific, for the orange line $v_B = \Delta t_d / \Delta x$ in Table (1), we have

$$1 = \int_0^{L'} \left|\psi'(t'=0, x')\right|^2 \mathrm{d}x' = \int_0^L A^2 \left|\psi(t=v_B x, x)\right|^2 \frac{\mathrm{d}x}{\gamma} = \frac{A^2}{\gamma} \frac{^0 n_0 + {}^1 n_1 + {}^2 n_2 + \cdots + {}^m n_m}{N} \tag{3.6}$$

by which the normalization constant $A$ can be fixed.

Coming back to Eq. (2.10), the Lorentz transformed wave-function can be given as

$$\psi'(t', x') = \int \phi'(k') e^{-i(\omega'_{Ek} t' - k' x')} \mathrm{d}k' \tag{3.7}$$

in which $\begin{cases} \omega'_{Ek} = \gamma(\omega_E - vk) \\ k' = \gamma(k - v\omega_E) \end{cases}$ is the Lorentz transformed off-shell energy-momentum vector $(\omega_E, k)$, and $\phi'(k') = \phi(k)$. Since $\mathrm{d}k' = \gamma \mathrm{d}k$ and $e^{-i(\omega'_{Ek}t'-k'x')} = e^{-i(\omega_E t - kx)}$, Eq. (3.7) and Eq. (2.10) satisfy Eq. (2.6). For constant eigen-energy $\omega_E$, the wave-function given by Eq. (2.10) is static. However, the wave-function given by Eq. (3.7) is no longer static in a sense that $\left| \psi'(t', x') \right|^2$ is time-dependent. Thus, for non-free massive particles, the Lorentz transformed static probability distribution becomes a time-dependent probability distribution.

## 4. Invariant of probability under other transformations

In Minkowski space-time, other coordinate transformations include: spatial translation and spatial rotation. For these transformations, the wave-function $\psi'(\vec{x}', t)$ in transformed $(\vec{x}', t)$ coordinate still satisfies $\psi'(\vec{x}', t) = \psi(\vec{x}, t)$, it can be justified by demanding the probability invariance viewed in transformed coordinates as given by the *postulate*. Note that for coordinate transformation in curved space-time, the *postulate* still holds but the Born rule needs to be modified, which will be investigated in our next work.

## 5. Discussion

For the position measurement of a particle, since the simultaneous measurements in one frame become measurements performed at different times in another frame, one needs to compare the probability distribution regarding the particle's position measurements at different space-time sectors. For a plane wave $e^{i(\omega t - \vec{k} \cdot \vec{x})}$ which is form invariant under Lorentz transformation, the transformed wave in $(t', x')$ is $e^{i(\omega' t' - \vec{k}' \cdot \vec{x}')}$, $(\omega, \vec{k})$ transforms as a Lorentz-vector in order to be consistent with our result obtained as Eq. (2.6), i.e., $e^{i(\omega' t' - \vec{k}' \cdot \vec{x}')} = e^{i(\omega t - \vec{k} \cdot \vec{x})}$. Note that Eq. (2.6) is obtained without considering the form of the potential; the transformation rule can be inferred from the physical meaning of the wave-function itself rather than the equations (such as Schrödinger equation or Dirac equation). As a result, Eq. (2.6) demands how the potential transforms by changing reference frames.

Therefore, one can see that the invariant of quantum probabilities in different reference frames can be treated as a most fundamental rule in nature which demands how potentials and physical quantities, such as the 4-momentum $(\omega, \vec{k})$, transform in different frames. This is in analogy with the one in classical physics: 4-velocity $(\frac{dt}{d\tau}, \frac{d\vec{x}}{d\tau})$ of an object transforms as a Lorentz-vector is to ensure that this object can always be found at that transformed space-time location with a definite probability 1. The transformation rule given by Eq. (2.6) is simple but fundamental, it is deduced from a basic *postulate*. It holds for any wave-function $\psi(t, x)$ of a single particle interacting with an arbitrary potential. This work sheds new light onto the inner workings of the quantum world.